\documentclass[aps,pra,twocolumn,superscriptaddress]{revtex4-2}
\usepackage[T1]{fontenc}
\usepackage{amsmath}
\usepackage{amssymb}
\usepackage{amsfonts}
\usepackage{mathtools}
\usepackage{braket}
\usepackage{hyperref}

\DeclareMathOperator\erf{erf}

\begin{document}


\title{Natural orbitals and their occupation numbers for free anyons in the magnetic gauge}



\author{Jerzy Cioslowski}
\affiliation{Institute of Physics, University of Szczecin, Wielkopolska 15, 70-451 Szczecin, Poland}
\affiliation{Max-Planck-Institut für Physik komplexer Systeme, Nöthnitzer Straße 38, 01187 Dresden, Germany}

\author{Oliver M. Brown}
\affiliation{School of Mathematics, University of Bristol, Fry Building, Woodland Road, Bristol, BS8 1UG, United Kingdom}

\author{Tomasz Maciazek}
\email{tomasz.maciazek@bristol.ac.uk}
\affiliation{School of Mathematics, University of Bristol, Fry Building, Woodland Road, Bristol, BS8 1UG, United Kingdom}

\date{\today}

\begin{abstract}
We investigate the properties of natural orbitals and their occupation numbers of the ground state of two non-interacting anyons characterised by the fractional statistics parameter $\alpha$ and confined in a harmonic trap. We work in the boson magnetic gauge where the anyons are modelled as composite bosons with magnetic flux quanta attached to their positions.
We derive an asymptotic form of the weakly occupied natural orbitals, and show that their corresponding (ordered descendingly) occupation numbers decay according to the power law $n^{-(4+2\alpha)}$, where $n$ is the index of the natural orbital. We find remarkable numerical agreement of the theory with the natural orbitals and their occupation numbers computed from the spectral decomposition of the system's wavefunction. We explain that the same results apply to the fermion magnetic gauge.
\end{abstract}

\maketitle

\section{Introduction}
Quantum statistics describes how the wavefunction of a multi-particle quantum system changes under particle exchange.
In three dimensions, particles can have statistics that are either fermionic or bosonic.
Anyons are class of particles that exist only in two dimensions with fractional statistics, between that of bosons and fermions, allowing their states to gain arbitrary complex phases under exchange \cite{Leinaas1977OnTT,Wilczek1982,Goldin80,Goldin85,Hansson}.
Such a statistics plays a role in understanding the fractional quantum Hall effect (FQHE), with emergent quasiparticles that have been identified as anyons \cite{ArovasSchriefferWilczek1984}. In this model, anyons are assumed to form a non-interacting gas (reviewed in Section \ref{sec:background} of this paper) whose free Hamiltonian is unitarily equivalent to a system composed of either fermions or hard-core bosons with magnetic flux quanta bound to their positions \cite{laughlin1988,Wu1984,Lundholm2016}. This picture is usually refereed to as the {\textit{magnetic gauge}}. The magnetic flux quanta realise the exchange phases as the Aharonov-Bohm phases. Anyons also find application in quantum computing, as the exchange of non-abelian anyons (described by multicomponent wavefunctions) allows the implementation of topologically protected quantum gates which are intrinsically robust against local noise \cite{nayaksimonreview2008,Pachos_2014,A_Yu_Kitaev_2001,KITAEV20032}. Such anyonic quasiparticles can be detected experimentally \cite{Feldman_2021} using interferometry \cite{BONDERSON,Nakamura2020,PhysRevX.13.041012}, correlation measurements in collider geometries \cite{doi:10.1126/science.aaz5601} or spectroscopy \cite{PhysRevX.8.011037}.

Due to the presence of anyonic quasiparticles in the FQHE, there has recently been interest in developing the Kohn-Sham density functional theory (KS-DFT, see \cite{HK64,KS65} for more background)  for the treatment of confined non-interacting many-anyon systems. In particular, KS-DFT has been developed for anyons described as composite flux-fermions \cite{HJ19,anyonsdftkohnsham}. The reduced density matrix functional theory (RDMFT \cite{Gilbert1975}) is suitable for describing strongly correlated systems \cite{PirisBenchmark}. As anyonic quasiparticles appear in strongly correlated systems, this method has been applied to an FQHE system \cite{ToloHarju2010} (specifically, to a confined quantum dot in the spin-frozen strong magnetic field regime). Recent developments in the foundations of RDMFT for bosons \cite{BWMS20} suggest that RDMFT might be a suitable tool also for studying anyon gas in the boson magnetic gauge.

In many methods of quantum chemistry the choice of the appropriate single-particle basis is a crucial step. For instance, the Hartree-Fock method approximates the $N$-body wavefunction of a system as a single Slater determinant. This approach has been applied to study the fermion-flux composites in relation to superconductivity and the presence of the energy gap in the free anyon gas  \cite{laughlin1988, hannalaughlin1989}. More generally, the multi-configurational self-consistent field method (MCSCF) relies on superposing several Slater determinants coming from a given single-particle basis. For a given $N$-particle (bosonic or fermionic) quantum state, there exists a distinguished multi-configurational expansion called the {\textit{natural expansion}}. The single-particle basis that is used in the natural expansion consists of the {\textit{natural orbitals} (NOs) that are defined as eigenfunctions of the state's one particle reduced density matrix (1RDM). It is known that the natural expansion has the fastest convergence of all the possible expansions \cite{lowdin1956}. The eigenvalues of the 1RDM are called the {\textit{natural occupation numbers}} (NONs) and denoted by $\nu_n$, $n=1,2,\dots$, where $\nu_1\geq\nu_2\geq\dots$. In this paper, we employ the convention for the 1RDM to be always of trace one. Due to normalisation, the NONs (ordered descendingly) have to decay to zero if the single-particle basis is of infinite dimension. The rate of their decay reflects certain fundamental features of the system at hand. For instance, only the first $N$ NONs of an uncorrelated state of $N$ electrons (a Slater determinant) are nonzero. On the other hand, a slow decay of NONs characterises a highly correlated quantum state. Furthermore, the knowledge of the NOs and NONs allows one to compute the expectation value of any one-particle observable where the individual contributions of the NOs are proportional to their corresponding NONs. For these fundamental reasons, the asymptotic rate of decay of NONs in different quantum systems has been an object of notable interest in quantum chemistry. 

One of the early results in this area is Hill's asymptotic \cite{Hill85} which concerns ground states of two-electron systems confined in an external potential with spherical symmetry. Due to this symmetry, the 1RDM has a block-diagonal structure where the blocks are labelled by the angular momentum quantum number $l$. Hill's asymptotic states that the total occupancies of the blocks, $\omega_l = \sum_{n} \nu_{nl}$, satisfy
\begin{equation}
\lim_{l\to\infty} \left(l+\frac{1}{2}\right)^7\omega_l = \mathcal{C}_H,
\end{equation}
where $\mathcal{C}_H$ is a constant that can be computed explicitly from the given quantum state. This asymptotic behaviour of $\omega_l$ was anticipated by the preceding numerical calculations for the helium atom \cite{Lakin65,CSM79}. Numerical calculations for the harmonium atom \cite{CB05,C15} further confirmed the validity of Hill's asymptotic beyond Coulomb external potentials. The large-$n$ asymptotic of $\nu_{nl}$ (for fixed $l$-sector) has proved to be a more difficult problem to study due to the lack of sufficiently accurate electronic structure data. Various conjectural forms of this asymptotic have been proposed and subsequently refuted over the years \cite{Bunge1970,CSM79,CP19}. Finally, its correct form has been determined for ground states of two-electron systems in central potentials and proved rigorously to be \cite{cioslowski2019}
\begin{equation}\label{jc_asymp}
\lim_{n\to\infty} n^8 \nu_{nl} = \mathcal{A},
\end{equation}
where the constant $\mathcal{A}$ does not depend on $l$ and can be calculated explicitly from the wavefunction at hand. This result has been subsequently generalised to singlet states of systems with arbitrary symmetries and numbers of electrons \cite{CS21}. Another work concerning general quantum systems of $N$ Coulomb-interacting particles proved the asymptotic \cite{sobolev}
\begin{equation}\label{sobolev_asymp}
\lim_{n\to\infty} n^{\frac{8}{3}} \nu_{n} = \mathcal{B},
\end{equation}
where the constant $\mathcal{B}$ can also be calculated explicitly from the wavefunction \cite{CS21,sobolev}. Note the lack of the subscript $l$ in the formula \eqref{sobolev_asymp} that does not assume any symmetries. The above results have subsequently led to the recent discovery of a universal power law governing the accuracy of wave function-based electronic structure calculations \cite{C22}. Similar results have been recently obtained for a system with the Fermi-Huang interparticle interaction \cite{CETH23}.

For the reliable application of quantum chemistry methods to anyonic systems it is necessary to understand similar asymptotic behaviour for models of non-interacting anyons. Here it is relevant that anyonic systems are two-dimensional as opposed to the previously mentioned quantum-chemical systems which are three-dimensional. Due to this change in dimensionality, new technical tools have to be applied in order to extend the methodology of the above cited papers to anyonic systems. However, the core feature remains true: the NOs and NONs are solutions to the eigenproblem of an integral operator whose kernel has a particle-particle coalescence cusp that drives the large-$n$ asymptotic of the NOs and NONs. More specifically, as we explain in Section \ref{sec:background}, the coalescence cusp is proportional to $|z_1-z_2|^\alpha$, where $z_1,\ z_2$ are the positions of the two boson-flux composite particles (represented by complex numbers) and $\alpha\in[0,1]$ is the fractional statistics parameter. In consequence, we show that the {\textit{natural amplitudes}} (NAs, the positive square roots of the NONs, $\sigma_{nl}^2=\nu_{nl}$) of the ground state of the system at hand satisfy
\begin{equation}
\lim_{n\to\infty} n^{\alpha + 2} \sigma_{nl} = \mathcal{D}(\alpha),
\end{equation}
and provide explicit expression for the constant $\mathcal{D}(\alpha)$ in Equation \eqref{eq:NA_asymptotics} in Section \ref{sec:asymptotics}. We also find very accurate asymptotic forms of the weakly occupied NOs in terms of appropriately transformed integer Bessel functions.


\section{Theoretical background}
\label{sec:background}
Anyons have quantum statistics between fermions and bosons -- the exchange of a pair of abelian anyons results with the multiplication of the many-anyon wavefunction by the phase factor $e^{i\pi \alpha}$, where $\alpha \in [0,1]$.
Consequently the wavefunction $\Psi_{\alpha}$ of $N$ abelian anyons can be expressed in terms of a bosonic wavefunction $\Psi_{B}$ as 
\begin{equation}\label{eq:statisticstrans}
    \Psi_{\alpha} (z_1,..., z_N) = \prod_{j<k}^{N} \frac{(z_j - z_k)^{\alpha}}{|{z_j - z_k}|^{\alpha}} \Psi_{B} (z_1,..., z_N),
\end{equation}
where the complex number $z_j = x_j + i y_j$ describes the position of the $j$-th particle, and $\alpha$ is the fractional statistics parameter (see e.g. \cite{laughlin1988, Douglas}). Similar mapping defines anyons in one dimension, where the one-body reduced density matrix of anyons confined in a harmonic trap and its NOs can be computed efficiently even for large particle numbers \cite{PhysRevA.93.063627}, showing that the largest NON follows a power law $\nu_1\sim N^{p(\alpha)}$ with $0<p(\alpha)<1$.
The anyonic wavefunction $\Psi_{\alpha}$ interpolates between bosons for $\alpha=0$ and fermions for $\alpha=1$.
The free-particle Hamiltonian $\hat H_{free} = \mathbf{\nabla}_{1}^2+\mathbf{\nabla}_{2}^2$ acting on $\Psi_{\alpha}$ then transforms under the gauge transformation in Equation\ \eqref{eq:statisticstrans} to a magnetic Hamiltonian acting on $\Psi_{B}$ \cite{Wilczek1982,hannalaughlin1989}.
In the case of two particles, this magnetic Hamiltonian in centre of mass coordinates can be expressed as \cite{wu1984harmonic}
\begin{equation}\label{eq:hamB}
    \hat{H}_{B} = - \frac{\hbar^{2}}{4 m} \mathbf{\nabla}_{Z}^2 + \frac{1}{m}\big( -i \hbar \mathbf{\nabla}_{z} - e \mathbf{A}(z) \big)^2
\end{equation}
where $Z$ is the centre of mass, $z = z_1 - z_2$ is the relative coordinate and $\mathbf{A}(z)$ is the magnetic vector potential
\begin{equation}
    \mathbf{A}(z) = \Phi \begin{pmatrix} - \textrm{Im}(z_1 - z_2) \\ \textrm{Re}(z_1 - z_2) \end{pmatrix} = \Phi {|z_1 - z_2|} \mathbf{\hat{e}}_{\theta},
    \end{equation}
where $e \Phi = \alpha \hbar$.
In this model, the anyonic exchange phases can be thought of as Aharonov-Bohm phases due to the presence of a magnetic vector potential in the relative Hamiltonian.
The bosonic Hamiltonian therefore describes boson-flux composites with hard cores.

The eigenstates for this two-anyon system in a harmonic potential are known \cite{wu1984harmonic}.
Under the potential $V_{external} = (z_1^2+z_2^2) / 2$ (here and in the following the "atomic" units are used in which the harmonic potential strength, the particle mass and $\hbar$ are all equal to one) the ground state wavefunction in the boson-flux composite picture is given by
\begin{equation}\label{psiB}
\begin{aligned}
    & \Psi_{B}^{(\alpha)} (z_1,z_2)  = \mathcal{N}_{\alpha}\, |z_1-z_2|^{\alpha} e^{-\frac{|z_1|^2+|z_2|^2}{2}},\\
& \mathcal{N}_\alpha=\frac{1}{\pi}\frac{1}{\sqrt{2^{\alpha}\Gamma(\alpha+1)}}. 
\end{aligned}
\end{equation}

The natural occupation numbers (NONs) $\nu_n$ for a many-particle quantum system are defined as the eigenvalues of the one-particle reduced density matrix (1RDM). Recall that for a two-particle wavefunction in $\mathbb{R}^d$ the 1RDM reads \cite{Lieb83}
\begin{equation}
    \gamma (\mathbf{r},\mathbf{r}') = \int_{\mathbb{R}^d} d \mathbf{r}_2 \Psi (\mathbf{r},\mathbf{r}_2) \overline{\Psi} (\mathbf{r}',\mathbf{r}_2),
\end{equation}
and the natural orbitals $\phi_{n}$ are its eigenstates, i.e.
\begin{equation}
    \int_{\mathbb{R}^d} d \mathbf{r'} \gamma (\mathbf{r},\mathbf{r}') \phi_n (\mathbf{r'}) = \nu_n \phi_n (\mathbf{r}).
\end{equation}

If the wavefunction $\Psi$ is real and symmetric, it is known that its NOs and NONs can be also found by solving the following homogeneous Fredholm equation of the second kind \cite{lowdinschull1956}, which circumvents the necessity of computing $\gamma$
\begin{equation}
    \int_{\mathbb{R}^d} d \mathbf{r}_2 \Psi (\mathbf{r},\mathbf{r}_2) \phi_n (\mathbf{r}_2) = \sigma_n \phi_n (\mathbf{r_1}).
\end{equation}
In the above equation the eigenvalues $\{\sigma_n\}_{n=1}^\infty$ are the natural amplitudes (NAs), $\nu_n=\sigma_n^2$.

Moreover, if the wavefunction $\Psi$ is rotationally symmetric (as is the case for any eigenfunction of the Hamiltonian \eqref{eq:hamB}), then the relevant eigenproblem is block-diagonal where the blocks are enumerated by the angular momentum quantum number $l$. In summary, the rest of this paper will be devoted to asymptotically solving the integral equation
\begin{equation}\label{eq:main}
\int_{\mathbb{R}^2} dz_2 \Psi_{B}^{(\alpha)}(z_1,z_2) \phi_{nl} (z_2) = \sigma_{nl} \phi_{nl} (z_1),
\end{equation}
where $\Psi_{B}^{(\alpha)}$ is given by formula \eqref{psiB}, for an arbitrary fixed spin sector $l$, in the limit of $n\to \infty$ which means studying the weakly occupied NOs and their corresponding NAs. The key intuitions here are that i) the radial part of $\phi_{nl}$ is highly oscillatory (as shown in Fig. \ref{fig:NOs}a-c -- the $n$th numerically computed NO has $n-1$ nodes), ii) when integrating a highly oscillatory function against a function that has discontinuous derivatives, the result goes to zero as a polynomial of the inverse of the oscillation frequency.

\section{Derivation of asymptotics}\label{sec:asymptotics}

Restricting to a fixed-$l$ sector means taking the NOs of the following forms
\begin{equation}\label{eq:natorbform}
    \phi_{nl}(z)=e^{i\theta l}\psi_{nl}(r)e^{-r^2/2},
\end{equation}
where the factor $e^{-r^2/2}$ is included for the sake of convenience. The orthonormality condition of the NOs then reads
\begin{gather}\label{eq:orthonormality}
    2\pi \int_0^\infty rdr\, \psi_{n_1l}(r)\psi_{n_2l}(r)e^{-r^2}=\delta_{n_1,n_2}.
\end{gather}
Using the relative angle variable $\theta_{12}=\theta_1-\theta_2$ and the polar coordinates $z_j = r_j e^{i\theta_j}$, $j=1,2$, the Fredholm Equation\ \eqref{eq:main} becomes
\begin{equation}\label{eq:fredholm_l}
\begin{aligned}
    & \mathcal{N}_\alpha \int_0^\infty r_2 dr_2 \int_0^{2\pi} d\theta_{12} \cos\left(\theta_{12}l\right) \times \\
    & \left(r_1^2+r_2^2-2 r_1 r_2 \cos\theta_{12}\right)^{\alpha/2}\, e^{-r_2^2}\psi_{nl}(r_2)=\sigma_n\,\psi_{nl}(r_1).
\end{aligned}
\end{equation}
In order to evaluate the angular integral above, note that the function $\cos\left(\theta_{12}l\right)$ can be written as a polynomial of degree $2l$ in the variable $t=\left(r_1^2+r_2^2-2 r_1 r_2 \cos(\theta_{12})\right)^{1/2}$.
This is because
\begin{equation}
\cos(\theta_{12}l) = T_l\left( \cos\theta_{12} \right) = T_l\left( \frac{r_1^2+r_2^2-t^2}{2 r_1 r_2} \right),
\end{equation}
where $T_l$ is the Chebyshev polynomial of order $l$, which then allows the finite polynomial expansion
\[\cos(\theta_{12}l) = \sum_{p=0}^{2l} a_{p}^{(l)}(r_1,r_2) t^{2p},\]
where the coefficients $a_{p}^{(l)}$ can be found using the explicit expansion formulas for $T_l$. In particular, 
\[a_{0}^{(l)}(r_1,r_2) = T_l\left(\frac{r_1^2+r_2^2}{2 r_1 r_2}\right).\]
Plugging this expansion into the LHS of\ \eqref{eq:fredholm_l}, we get
\begin{equation}\label{eq:chebyshev_expansion}
\begin{aligned}
   & \mathcal{N}_\alpha \sum_{p=0}^{2l}\int_0^\infty r_2 dr_2\, a_{p}^{(l)}(r_1,r_2) e^{-r_2^2}\psi_{nl}(r_2) \times \\
    &\times \int_0^{2\pi} d\theta_{12} \left(r_1^2+r_2^2-2 r_1 r_2 \cos\theta_{12}\right)^{\alpha/2+p} = \\
   & = \sigma_{nl}\,\psi_{nl}(r_1).
\end{aligned}
\end{equation}
Next, we use the result that for any $\beta\in\mathbb{R}$,
\begin{equation}\label{eq:theta12_integral}
\int_0^{2\pi} d\theta_{12} \left(r_1^2+r_2^2-2 r_1 r_2 \cos\theta_{12}\right)^{\frac{\beta}{2}} = \pi\, G_\beta(r_1,r_2),
\end{equation}
where
\begin{equation} \label{eq:I_beta}
\begin{aligned}
& G_\beta(r_1,r_2)=| r_1-r_2| ^\beta \, _2F_1\left(\frac{1}{2},-\frac{\beta}{2};1;\frac{-4 r_1 r_2}{(r_1-r_2)^2}\right) \\
& +(r_1+r_2)^\beta \, _2F_1\left(\frac{1}{2},-\frac{\beta}{2};1;\frac{4 r_1 r_2}{(r_1+r_2)^2}\right),
\end{aligned}
\end{equation}
and $_2F_1$ is the ordinary hypergeometric function. 

So far, all the calculations have been exact. However, we will next start making approximations in order to make the large-$n$ asymptotic solution to the Equation \eqref{eq:fredholm_l} more tractable.
The key fact to notice is that the leading contribution to the $n\to \infty$ asymptotic of $\sigma_n$ comes from the lowest order cusp of the LHS of Equation \eqref{eq:fredholm_l} around $r_1=r_2$.
Let us next look at the asymptotic expansion of $G_\beta(r_1,r_2)$ around $r_1=r_2$ when $0<\beta<1$.
The only contribution to the cusp around $r_1=r_2$ is the first term in the RHS of Equation \eqref{eq:I_beta}.
Its asymptotic expansion reads
\begin{equation}\label{eq:I_asymp}
\begin{aligned}
&\tilde G_\beta(r_1,r_2)=|r_1-r_2| ^\beta \, _2F_1\left(\frac{1}{2},-\frac{\beta}{2};1;\frac{-4 r_1 r_2}{(r_1-r_2)^2}\right)  \\
 &   = \frac{2^\beta\, \Gamma\left(\frac{1+\beta}{2}\right)}{\sqrt{\pi}\Gamma\left(1+\frac{\beta}{2}\right)}\, r_1^\beta - \frac{\beta\, 2^{\beta-1}\, \Gamma\left(\frac{1+\beta}{2}\right)}{\sqrt{\pi}\,\Gamma\left(1+\frac{\beta}{2}\right)}\, r_1^{\beta-1} (r_1-r_2)  \\
 &   + \frac{\Gamma\left(-\frac{1+\beta}{2}\right)}{2\sqrt{\pi}\,\Gamma\left(-\frac{\beta}{2}\right)}\, \frac{1}{r_1} | r_1-r_2| ^{\beta+1} + \mathcal{O}((r_1-r_2)^{2}). 
   \end{aligned}
\end{equation}
The first three terms of this expansion are plotted in Fig. \ref{fig:asymptotic}. In the above formula, we can see that the leading-order cusp is $|r_1-r_2| ^{\beta+1}$. Similarly, when $\beta>1$ the analogous cusp will appear in the order $\beta+1$.
This in turn means that the leading-order cusp in the Equation \eqref{eq:chebyshev_expansion} will come from the ($p=0$)-summand. Thus, the large-$n$ asymptotic solution to the Equation\ \eqref{eq:fredholm_l} can be obtained from the following equation which extracts only the leading-order cusp around $r_1=r_2$.

\begin{figure}
    \centering
    \includegraphics[width=0.45\textwidth]{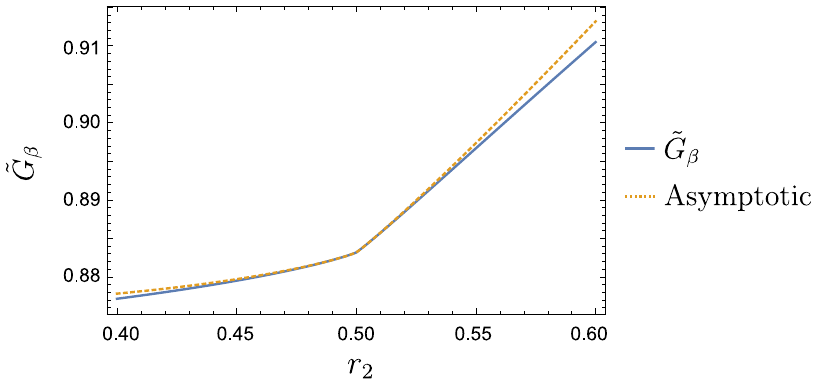}
    \caption{The leading-order cusp in the function $\tilde G_\beta(r_1,r_2)$ is reproduced by the first three terms of its asymptotic expansion around $r_1=r_2$. The plot shows the function $\tilde G_\beta(r_1,r_2)$ vs. the first three terms of its leading-order asymptotic expansion \eqref{eq:I_asymp} for $\beta=1/5$, $r_1=0.5$.}
    \label{fig:asymptotic}
\end{figure}

\begin{equation}\label{eq:I_asymp_cusp}
\begin{aligned}
& \mathcal{B}(\alpha)\,\int_0^\infty r_2 dr_2\,| r_1-r_2| ^{\alpha+1}e^{-r_2^2}\,\psi_{nl}(r_2)=\sigma_{nl}\,r_1\psi_{nl}(r_1), \\
& \mathcal{B}(\alpha)=\sqrt{\frac{1}{2^{\alpha}\, \pi\, \Gamma(\alpha+1)}}\, \frac{\Gamma\left(-\frac{1+\alpha}{2}\right)}{2\,\Gamma\left(-\frac{\alpha}{2}\right)}.
\end{aligned}
\end{equation}
In the above equation we have also used the fact that $a_0^{(l)}(r_1,r_1)=1$.

In analogy to the three-dimensional case \cite{cioslowski2019} we choose an ansatz for $\psi_{n,l}$ of the form
\begin{equation}\label{eq:phiansatz}
    \psi_{nl}(r)=f_{nl}(r)\,J_{l}\left(\kappa_{nl} g_{nl}(r)\right),
\end{equation}
where
\begin{gather*}
    f_{nl}(r),\,g_{nl}(r)> 0,\quad g_{nl}'(r)> 0,\quad \kappa_{nl}>0,\\
    \infty>\lim_{r\to 0}f_{nl}(r)> 0,\quad \infty>\lim_{r\to 0}\frac{g_{nl}(r)}{r}> 0.
\end{gather*}
Functions $J_l$ are the Bessel functions of the first kind.
As previously mentioned, taking the limit $n\to \infty$ means considering highly oscillatory NOs, thus we necessarily have $\kappa_{nl}>>1$. The strategy is now to extract the leading order expansion of the Equation \eqref{eq:I_asymp_cusp} in the powers of $1/\kappa_{nl}$ which will lead us to certain consistency condition for the function $g_{nl}(r)$. Because $\kappa_{nl}$ is large, we can approximate the Bessel functions as \cite{abramowitz1965handbook}
\begin{equation}
    J_l (z) \approx \sqrt{\frac{2}{\pi z}} \cos \left(z - (2l+1)\frac{\pi}{4}\right).
\end{equation}
For convenience, we will ignore the phase shift proportional to $\pi/4$ under the cosine, as it will not alter the resulting consistency relations.
Consequently, the integral equation \eqref{eq:I_asymp_cusp} becomes
\begin{gather}
    \mathcal{B}(\alpha)\int_0^{\infty} dr_2\, | r_1-r_2|^{\alpha+1}\, \frac{r_2\ e^{-r_2^2} f_{nl}(r_2)}{\sqrt{g_{nl}(r_2)}} \cos\left(\kappa_{nl} g_{nl}(r_2)\right)\nonumber\\
    = \sigma_{nl}\, r_1\, \frac{f_{nl}(r_1)}{\sqrt{g_{nl}(r_1)}} \cos\left(\kappa_{nl} g_{nl}(r_1)\right)\label{eq:fredholminteqcos}.
\end{gather}

With the change of variables \(u_i := g_{nl}(r_i) \quad i=1,2\), \(q(u) := g_{nl}^{-1}(u)\), we define
\begin{equation*}\label{eq:fn}
    F (u_2) = \frac{1}{\sqrt{u_2}} q(u_2)\, q'(u_2)\, e^{-\left(q(u_2)\right)^2} f_{nl}(q(u_2)).
\end{equation*}
Next, we extract the leading asymptotic around $u_1=u_2$ using the fact that
\begin{gather*}
    \left|q(u_1)-q(u_2)\right|^{\alpha+1}=\left|u_1-u_2\right|^{\alpha+1} (q'(u_1))^{\alpha+1} + \\
    +\, \mathcal{O}(|u_1-u_2|^{\alpha+2}).
\end{gather*}
This allows us to write the integral Equation\ \eqref{eq:fredholminteqcos} as
\begin{equation}\label{eq:integral_real_F}
\begin{aligned}
   & \mathcal{B}(\alpha)\,\Re \left\{ \int_0^{u_\infty} du_2\, F (u_2) |u_1-u_2|^{\alpha+1}e^{i\kappa_{nl} u_2} \right\} \\
&    = \sigma_{nl}\, \frac{q(u_1) f_{nl}(q(u_1))}{\sqrt{u_1} (q'(u_1))^{\alpha+1}} \cos\left(\kappa_{nl} u_1\right),
\end{aligned}
\end{equation}
where $u_\infty = \lim_{r\to\infty}g_{nl}(r)$. Finally, we use a theorem from the paper \cite{kangshao2014} to extract the leading order behaviour when in $\kappa_{nl}\to\infty$ (see Appendix \ref{appendixA} for more details). This leads to
\begin{equation}\label{eq:fredholm_asymp_reduced}
\begin{aligned}
    &- \frac{2}{\kappa_{nl}^{\alpha+2}} \cos\left(\frac{\pi\alpha}{2}\right)\, \cos(\kappa_{nl} u_1) F\left(u_1\right) \Gamma{(\alpha + 2)}\\
    &= \sigma_{nl}\, q(u_1)\, \frac{q(u_1) f_{nl}(q(u_1))}{\mathcal{B}(\alpha) \sqrt{u_1} (q'(u_1))^{\alpha+1}} \cos \left(\kappa_{nl} u_1\right)
\end{aligned}
\end{equation}
Equating coefficients of \(\cos\left(\kappa_{nl} u_1\right)\) and expressing everything back in terms of the variable $r$, we arrive at the following consistency condition defining $g_{nl}(r)$
\begin{equation}\label{eq:kappa_g}
\begin{aligned}
 &   \kappa_{nl}\, g_{nl}(r) = \left(-\frac{\mathcal{C}(\alpha)}{\sigma_{nl}}\right)^{\frac{1}{\alpha+2}} \int_0^r dr_1\, e^{-r_1^2/(\alpha+2)} \\
    &= \left(-\frac{\mathcal{C}(\alpha)}{\sigma_{nl}}\right)^{\frac{1}{\alpha+2}} \mathcal{I}_\infty(\alpha) \erf \left(\frac{r}{\sqrt{\alpha+2}}\right),
\end{aligned}
\end{equation}
where erf is the error function, 
\begin{equation}
\mathcal{I}_\infty(\alpha) = \int_0^\infty dr_1\, e^{-r_1^2/(\alpha+2)} =  \frac{\sqrt{\pi (\alpha+2)}}{2}
\end{equation}
and
\begin{equation}
    \mathcal{C}(\alpha) = 2\mathcal{B}(\alpha) \Gamma(\alpha + 2) \cos \left(\frac{\pi \alpha}{2}\right).
\end{equation}
In order to determine the dependence of $\sigma_{nl}$ and $\kappa_{nl}$ on $n$ together with the form of the function $f_{nl}$, we refer to the orthonormality condition of the NOs \eqref{eq:orthonormality} which takes the explicit form
\begin{equation}\label{eq:NO_ortho_2}
\begin{aligned}
    2 \pi \int_{0}^{\infty} r dr f_{n_1 l}(r) f_{n_2 l}(r) J_{l}\left(\kappa_{n_1} g_{n_1 l}(r)\right) \times \\ 
    J_{l}\left(\kappa_{n_2} g_{n_2l}(r)\right) e^{- r^2} = \delta_{n_1,n_2}.
\end{aligned}
\end{equation}
In order to satisfy the above equality, we will make use of the orthogonality relation for Bessel functions \cite{abramowitz1965handbook}
\begin{equation}\label{bessel_ortho}
    \int_0^{1} dz\, J_l\left(\mu_{n_1} z\right) J_l\left(\mu_{n_2} z\right) z=  \frac{1}{2} J_l' (\mu_{n_1})^2\, \delta_{n_1,n_2},
\end{equation}
where $\mu_n$ is the $n$th node of $J_l$. Note first that if we require the relation
\begin{equation}\label{eq:f_const}
    \sqrt{\frac{r e^{- r^2}}{g_{nl}(r)\, g_{nl}'(r)}} f_{n l}(r) \equiv \mathcal{C}_{nl},
\end{equation}
to be satisfied and requiring that $g_{n_1 l}(r) = g_{n_2 l}(r)\equiv g_{l}(r)$ we can readily apply the identity \eqref{bessel_ortho} to the LHS of \eqref{eq:NO_ortho_2} after the familiar change of variables $u = g_l(r)$ under the integral. Additionally, recall that the $n$th NO must have $n-1$ nodes. In light of Equation \eqref{eq:kappa_g}}, this requires identifying the functions $g_{nl}(r)$ and $\kappa_{nl}$ as
\begin{equation}
g_{nl}(r)\equiv g(r) = \erf \left(\frac{r}{\sqrt{\alpha+2}}\right),\quad \kappa_{nl} = \mu_n.
\end{equation}
Consequently, we get that 
\begin{equation}
\left(-\frac{\mathcal{C}(\alpha)}{\sigma_{nl}}\right)^{\frac{1}{\alpha+2}} \mathcal{I}_\infty(\alpha)  = \mu_n.
\end{equation}
From the relation \eqref{eq:f_const} we determine that $f_{nl}(r)$ takes the form
\begin{equation}
    f_{nl}(r) =\mathcal{C}_{nl}\,  \sqrt{\frac{\erf{\frac{r}{\sqrt{\alpha+2}}}}{r}} e^{-\frac{r^2}{2(\alpha+2)}}e^{\frac{r^2}{2}} ,
\end{equation}
where the normalization factor reads
\begin{equation}
\mathcal{C}_{nl}=\frac{1}{|J_l'(\mu_n)|} \sqrt{\frac{2}{\pi \sqrt{\pi (\alpha+2)}}}.
\end{equation}
Using the well known asymptotic for the nodes of the Bessel functions $\mu_{n} \approx n \pi$, we summarize the results of this section as follows.
\begin{equation}\label{eq:NA_asymptotics}
\begin{aligned}
    &\lim_{n\to\infty}|\sigma_{nl}|\, n^{\alpha+2}= \mathcal{D}(\alpha),\\
    &\mathcal{D}(\alpha) = \frac{(2 \pi)^{-\alpha /2} (\alpha +2)^{\frac{\alpha }{2}+1} }{\sin \left(\frac{\pi  \alpha }{2}\right)\Gamma \left(-\frac{\alpha }{2}\right)^2 \sqrt{\Gamma (\alpha +1)}}.
\end{aligned}
\end{equation}

\begin{equation}\label{eq:NO_asymptotics}
\begin{aligned}
\phi_{nl}(r,\theta) \xrightarrow{n\to\infty} \mathcal{C}_{nl} \sqrt{\frac{\erf{\frac{r}{\sqrt{\alpha+2}}}}{r}} e^{-\frac{r^2}{2(\alpha+2)}} \times \\ 
\times J_l\left(\mu_n \erf \left(\frac{r}{\sqrt{\alpha+2}}\right)\right) e^{i\theta l}.
\end{aligned}
\end{equation}

\section{Numerical Methods}
In this section, we carry out the numerical verification of the asymototics predicted in Equations \eqref{eq:NA_asymptotics} and \eqref{eq:NO_asymptotics}. This is done by writing the Equation \eqref{eq:main} in the ($\kappa$-scaled) harmonic oscillator eigenbasis
\begin{equation}
\begin{aligned}
& \tilde \phi_{ml}^{(\kappa)}(r,\theta) = \mathcal{N}_{ml}^{(\kappa)}e^{i\theta l} (\kappa r)^{|l|} L_m^{|l|}\left(\kappa^2r^2\right)e^{-\kappa^2r^2/2}, \\
& \mathcal{N}_{ml}^{(\kappa)}=\kappa \sqrt{\frac{m!}{\pi (m+|l|)!}},
\end{aligned}
\end{equation}
where $\kappa>0$ and $L_m^{|l|}$ is a generalised Laguerre polynomial. These calculations are specific to $l=0$. As will be explained later, the parameter $\kappa$ will be optimised which allows us to pick the the basis in which the NONs converge at the fastest rate. To simplify the notation we  omit the $l$-subscripts, i.e. we write $\tilde \phi_{m,0}^{(\kappa)} \equiv \tilde \phi_{m}^{(\kappa)}$.

In the truncated basis $\phi_{m}^{(\kappa)}$, $m=0, 1,\dots, M$ the Equation \eqref{eq:main} for $l=0$ is solved via the diagonalisation of the $(M+1)\times(M+1)$ matrix
\begin{equation}\label{eq:A_entries}
\left[A^{(\alpha)}(\kappa)\right]_{m_1,m_2} =\Braket{\phi_{m_1}^{(\kappa)}(z_1)\phi_{m_2}^{(\kappa)}(z_2)|\Psi_B^{(\alpha)}(z_1,z_2)}.
\end{equation}
Let us next briefly describe the steps that we apply in order to evaluate some of the multidimensional integrals in the LHS of Equation \eqref{eq:A_entries}. Similarly to the methodology of Section \ref{sec:asymptotics}, we use the relative angle $\theta_{12}$ and compute the corresponding integral over $\theta_{12}$ using the result from Equation \eqref{eq:theta12_integral}. Next, we change the the radial coordinates $r_1$, $r_2$ to $r_1 = r \cos\xi$ and $r_2 = r \sin\xi$, $0\leq r\leq\infty$, $\xi\in[0,\pi/2]$. The Jacobian of this transformation is $r$. Under this change of variables, we have
\begin{equation}
\pi\, G_\alpha(r\cos\xi,r\sin\xi) = r^\alpha K_\alpha(\xi),
\end{equation}
where the function $K_\alpha(\xi)$ reads (after using an identity for the hypergeometric function to simplify its form)
\begin{gather*}
K_\alpha(\xi) = -\frac{2\sqrt{\pi}\,\Gamma\left(\frac{2+\alpha}{2}\right)}{\Gamma\left(\frac{3+\alpha}{2}\right)}\,\frac{\left(1+\sin(2\xi)\right)^{\frac{2+\alpha}{2}}}{|\cos(2\xi)|} \times  \\
\times\Im\left\{ _2F_1\left(\frac{2+\alpha}{2},\frac{1}{2},\frac{3+\alpha}{2},\left(\frac{1+\sin(2\xi)}{\cos(2\xi)}\right)^2\right)\right\}.
\end{gather*}
Next, we apply the polynomial expansions of the Laguerre polynomials. It turns out that the integrals over $r$ can be computed analytically in terms of the Euler gamma functions. Thus, we are only left with the task of evaluating numerically the integrals over $\xi$. After the above transformations, the matrix elements read

\begin{widetext}
\begin{equation}\label{eq:matrix_long}
\begin{aligned}
\left[A^{(\alpha)}(\kappa)\right]_{m_1,m_2} =\frac{\mathcal{N}_\alpha}{2\kappa^{\alpha/2}}\sum_{a=0}^{m_1}\sum_{b=0}^{m_2}\binom{m_1}{a}\binom{m_2}{b}\frac{(-1)^{a+b}}{a!b!} \Bigg(\frac{2 \kappa}{\kappa^2 + 1} \Bigg)^{2+a+b+\frac{\alpha}{2}} \Gamma \left(2+a+b+\frac{\alpha}{2}\right) \times \\
  \times  \left(\sum_{j=0}^a(-1)^j\binom{a}{j}\mathcal{J}_{b+j}^{(\alpha)}+\sum_{j=0}^b(-1)^j\binom{b}{j}\mathcal{J}_{a+j}^{(\alpha)}\right),
\end{aligned}
\end{equation}
\end{widetext}
where the integration over $\xi$ appears only in the integrals
\begin{equation}
\mathcal{J}_{k}^{(\alpha)} = \int_0^{\pi/4} d\xi\, \sin(2\xi) K_\alpha(\xi)\cos^{2k}(\xi).
\end{equation}
Note that in order to evaluate the expressions \eqref{eq:matrix_long} numerically for all $0\leq m_1,m_2\leq M$, we only need the integrals $\mathcal{J}_{k}^{(\alpha)}$ for $k=0,1,\dots,2M$ which can be pre-calculated separately. The difficulty of this approach is that we need to know the values of the integrals $\mathcal{J}_{k}^{(\alpha)}$ with very high precision, because they enter the alternating sums in Equation \eqref{eq:matrix_long}. To this end, we have used $Python$'s library $mpmath$. For the calculations presented in this section, we have set the size of the one-particle basis to $M=400$ and the precision to $2M$. Note also that the expressions \eqref{eq:matrix_long} can be computed efficiently for all $0\leq m_1,m_2\leq M$ at once using the vectorisation technique, i.e. by recognising that the Equation \eqref{eq:matrix_long} allows one to express matrix $A^{(\alpha)}(\kappa)$ as a product of three matrices.

The optimal paramater $\kappa$ has been chosen by maximising the fidelity
\begin{equation}
\mathcal{F}_M^{(\alpha)}(\kappa) = \sum_{m_1,m_2=0}^M \big{|}A_{m_1,m_2}^{(\alpha)}(\kappa)\big{|}^2.
\end{equation}
It has turned out that the optimal value of $\kappa$ is approximately independent of $\alpha$ and for $M=400$ it is $\kappa_{opt} \approx 7$.

\begin{figure}
    \centering
    \includegraphics[width=0.5\textwidth]{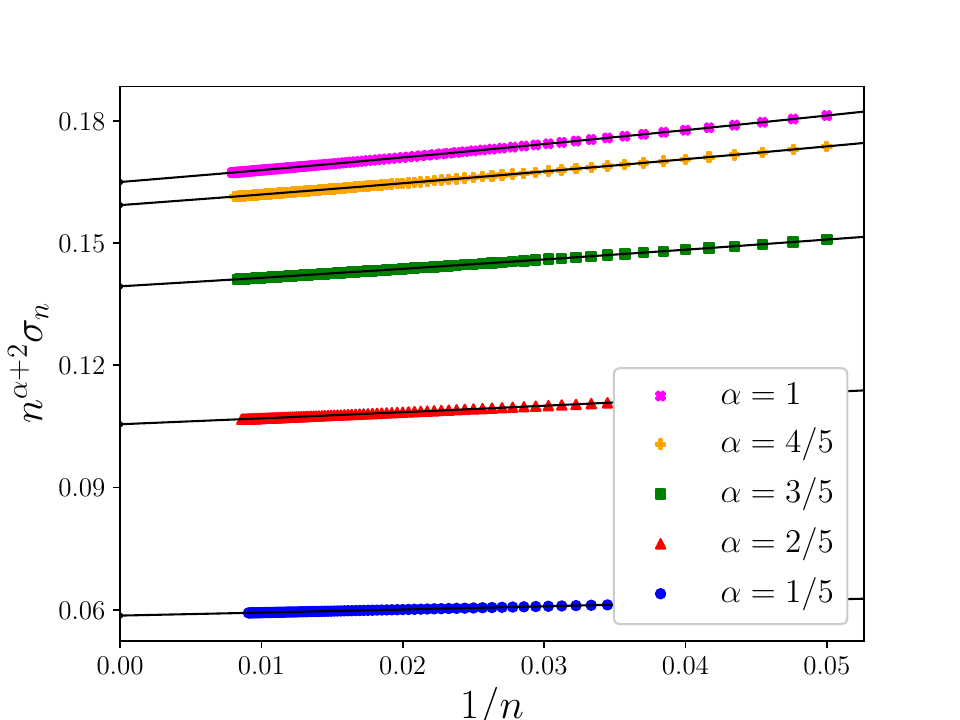}
    \caption{The NAs of the $l=0$ sector for $\alpha\in\left\{\frac{1}{5},\frac{2}{5},\frac{3}{5},\frac{4}{5},1\right\}$ and $M=400$.
    The number $n$ represents the index of the values $\sigma_{n}$. The solid lines show the outcome of the fifth order polynomial regression for each value of $\alpha$.}
    \label{fig:NAs_l0}
\end{figure}

The resulting NAs are plotted in Fig. \ref{fig:NAs_l0}. Thanks to the optimal choice of the parameter $\kappa$, we have obtained the convergence of the first $105$ - $125$ highest NAs (the exact number depends on $\alpha$) with the single-particle basis of the size $M=400$. In Fig. \ref{fig:NAs_l0} we have plotted the values of $n^{\alpha+2}\,\sigma_n$ vs. $1/n$, which allowed us to determine the values of the constants $\mathcal{D}(\alpha)$ with the relative accuracy of the order $10^{-3}$. As shown in Table \ref{tab:table_C}, this is in perfect agreement with the theoretical values calculated from Equation \eqref{eq:NA_asymptotics}.

\begin{table}
\begin{center}
  \begin{tabular}{| c || c |c |}
    \hline
    $\alpha$ & $\mathrm{Fitted\ }\mathcal{D}(\alpha)$ &  $\mathrm{Exact\ }\mathcal{D}(\alpha)$  \\ \hline \hline
    $1/5$ & $0.0586 \pm 0.0004$ & $0.058580\hdots$  \\ \hline
    $2/5$ & $0.1055 \pm 0.0005$ &  $0.105525\hdots$ \\ \hline
    $3/5$ & $0.1394 \pm 0.0005$ & $0.139367\hdots$  \\ \hline
    $4/5$ & $0.1593 \pm 0.0005$ & $0.159293\hdots$  \\ \hline
    $1$ & $0.1649 \pm 0.0005$ & $0.164961\hdots$  \\
    \hline
  \end{tabular}
\end{center}
 \caption{Comparison of the values of the constant $\mathcal{D}(\alpha)$ computed via the fifth-order polynomial regression in Figure \ref{fig:NAs_l0} for $l=0$ with the exact formula from Equation \eqref{eq:NA_asymptotics}.}
  \label{tab:table_C}
\end{table}

\begin{figure}
\centering
\includegraphics[width=0.5\textwidth]{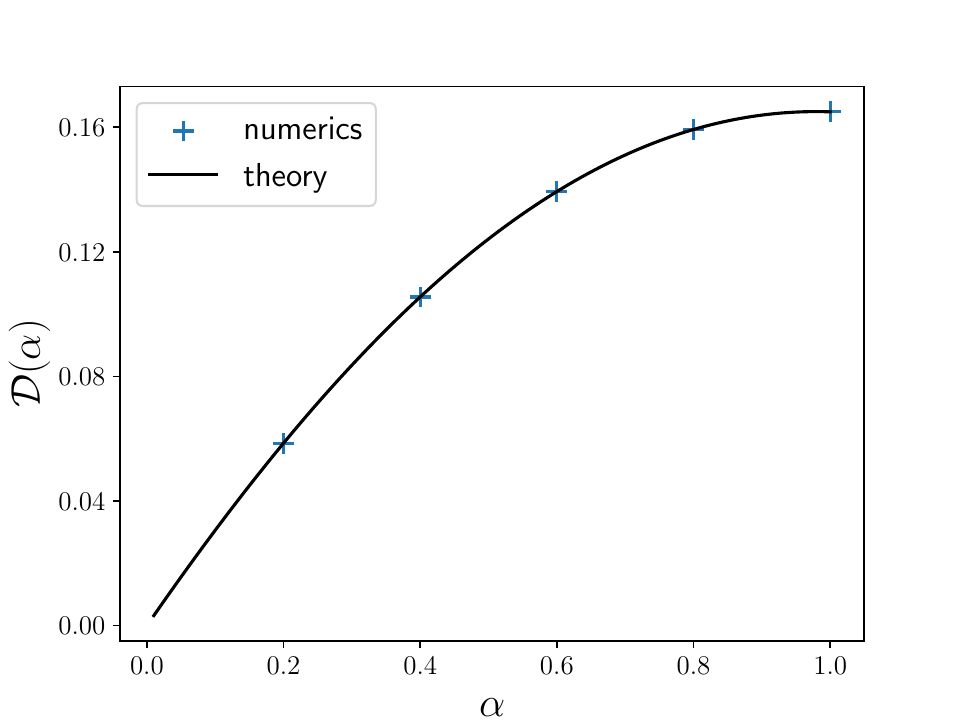}
\caption{The fitted values of $\mathcal{D}(\alpha)$ from fifth order polynomial regression for $\alpha\in\left\{\frac{1}{5},\frac{2}{5},\frac{3}{5},\frac{4}{5},1\right\}$ and $M=400$.
They are in perfect agreement with the formula\ \eqref{eq:NA_asymptotics}.}
\label{fig:a1l0}
\end{figure}

\begin{figure*}
    \centering
    \includegraphics[width=\linewidth]{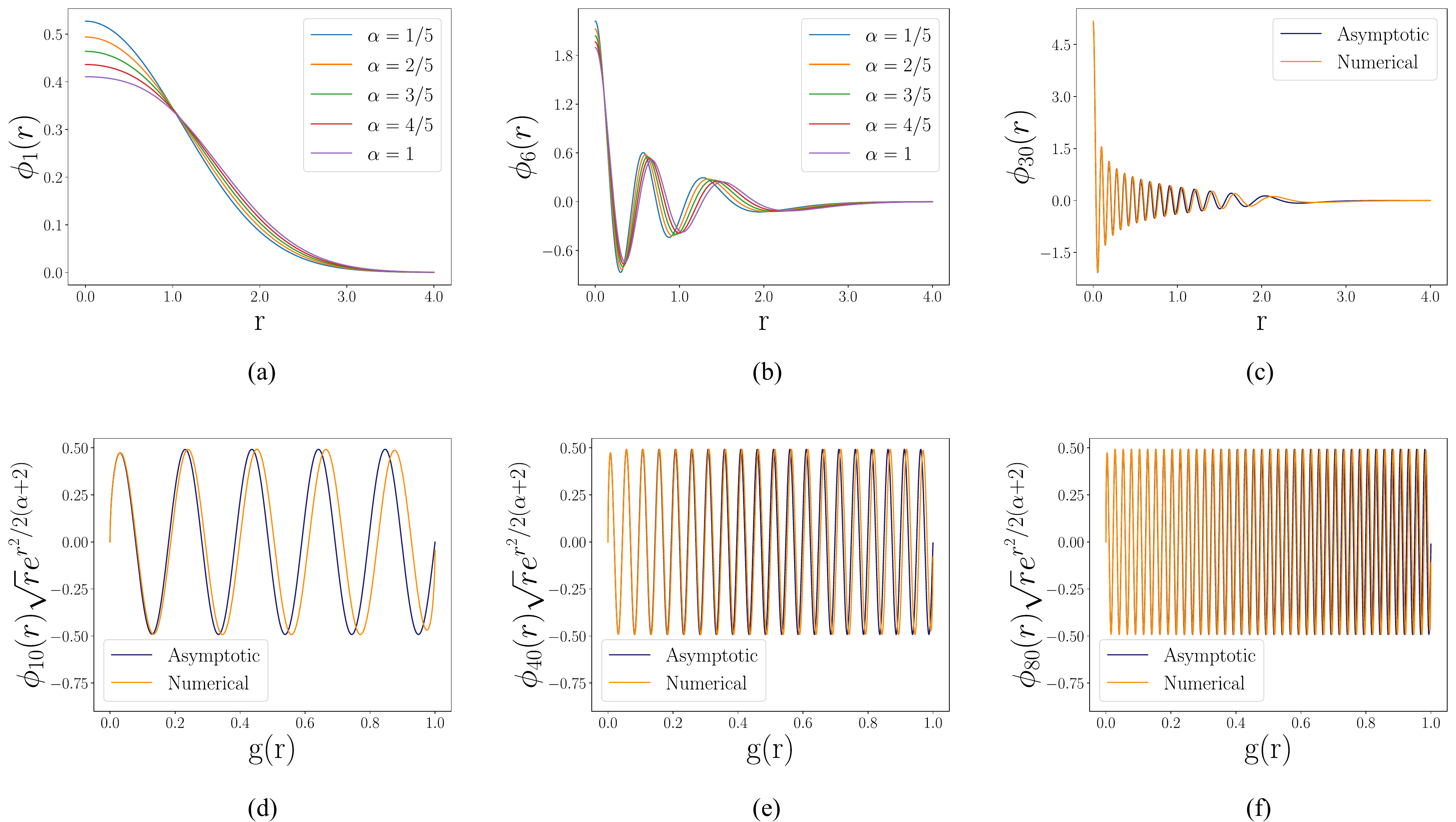}
    \caption{Top row: the $l=0$ natural orbitals of index 1 (a), and 6 (b) for various values of the fractional statistics parameter $\alpha$.
    Frame (c) shows a comparison of the numerical and asymptotic forms of the natural orbital $n=30$, for which $l=0$ and $\alpha = 1/5$.
    The natural orbitals are observed to cross the axis $n$ times, where $n$ is the principal quantum number of the orbital.
    Bottom row: a comparison of the numerically computed natural orbitals to their corresponding asymptotic formulae for large $n$, all with $l=0$ and $\alpha = 1/5$.
    For small $r$ the deviations negligible.
    Frames (d), (e) and (f) show the oscillatory component of the natural orbitals, with $n=10$ (d), $n=40$ (e) and $n=80$ (f).
    For small $n$, the deviations from the asymptotic formula diverge as $r$ increases from $0$.
    This divergence is less pronounced for larger $n$.}
    \label{fig:NOs}
\end{figure*}

In Fig. \ref{fig:NOs} we have plotted the numerically calculated NOs and compared them with their asymptotic forms from Equation \eqref{eq:NO_asymptotics}. We observe remarkable agreement of the asymptotic form even for values of $n$ as low as $30$. For the convenience of comparison, in the bottom row of Fig. \ref{fig:NOs} we have plotted the values of $\phi_n(r) \sqrt{r} e^{r^2/(2(\alpha+2))}$ vs. $g(r)$ which extracts the oscillatory part of the NOs. One can see that the NOs converge to their respective asymptotic forms very fast.

\section{Consequences of the NO and NA asymptotics for computing anyon correlations}

The asymptotic properties of NOs and NAs derived in the preceding sections are of broad relevance to physics of anyonic systems astthey show universal properties of anyonic wavefunctions. In particular, our results show that when an anyonic wavefunction is expanded in a single-particle basis, the convergence of such an expansion is much slower than in the case of fermionic or bosonic systems. This has fundamental consequences for applications of any numerical methods to anyonic systems. These consequences are particularly evident in two-particle systems where the natural orbitals and natural occupation numbers allow one to reconstruct the two-anyon wavefunction (up to a diagonal rotation of the single-particle basis), making it possible to compute the expectation value of any quantum observable. Such two-anyon systems have been extensively studied from the point of view of anyon correlations and anyon interferometry \cite{PhysRevLett.99.190401,PhysRevB.77.115442,PhysRevB.81.201306,PhysRevB.83.155440,PhysRevLett.116.156802,Subramanyan_2019} (including recent experimental proofs of anyonic statistics \cite{Nakamura2020, doi:10.1126/science.aaz5601}), where one of the relevant observables is the two-body observable called the bunching parameter \cite{Subramanyan_2019}.

In this section, we show in detail how to calculate such expectation values using the example of two-particle correlation coefficient $\tau$ defined as
\begin{equation}\label{eq:tau}
\tau = \frac{2\langle r_1 r_2\cos\theta_{12}\rangle}{\langle r_1^2+r_2^2\rangle},
\end{equation}
where we parametrise the configuration space of the anyons via $z_j = r_j e^{i\theta_j}$, $j=1,2$ with $\theta_{12}=\theta_1-\theta_2$ being the angle between the position vectors of the anyons. The correlation coefficient $\tau$ has been originally introduced in the context of electronic wavefunctions \cite{PhysRev.172.49}, however it is closely related to the anyon bunching parameter introduced more recently \cite{Subramanyan_2019}.

In order to calculate the correlation coefficient $\tau$ for the two-anyon system at hand, we will work in the boson magnetic gauge, i.e. use the wavefunction $\Psi_B^{(\alpha)}$ defined in \eqref{psiB}. It is convenient to change the variables to the centre of mass, $Z$, and relative position, $z$, given by
\begin{equation}\label{eq:cm_coords}
Z=\frac{1}{2}\left(z_1+z_2\right),\quad z = z_1 - z_2.
\end{equation}
In the new coordinates, the two-anyon wavefunction has the product form
\begin{equation}\label{eq:PsiB_cm}
\Psi_B^{(\alpha)}(z_1,z_2) = \mathcal{N}_\alpha e^{-|Z|^2} |z|^\alpha e^{-|z|^2/4}.
\end{equation}
What is more, the correlation coefficient also takes a simple form in the new coordinates, i.e.
\begin{equation}\label{eq:tau_cm}
\tau = \frac{4\langle |Z|^2\rangle-\langle |z|^2\rangle}{4\langle |Z|^2\rangle+\langle |z|^2\rangle}.
\end{equation}
It is a matter of a straightforward calculation to find the relevant expectation values. The result reads 
\begin{equation}\label{eq:tau_final}
\tau = -\frac{\alpha}{2+\alpha}.
\end{equation}
Recall that for $\alpha=0$ the quantum state $\Psi_B^{(\alpha)}$ is just a product state of two bosons which is uncorrelated. Consequently, the correlation coefficient vanishes for $\alpha=0$. What is more, $\tau$ is a monotonic function of the statistics parameter $\alpha$ and it reaches its minimum value $\tau=-1/3$ for $\alpha=1$, i.e. when the wavefunction $\Psi_B^{(\alpha)}$ describes a strongly correlated state of two hard core bosons.

Let us next take a closer look at how to compute the above correlation coefficient $\tau$ with the NOs of $\Psi_B^{(\alpha)}$. The two-anyon wavefunction can be written as
\begin{equation}
\Psi_B^{(\alpha)} (z_1,z_2) = \sum_{l=0}^\infty \sum_{n=0}^\infty c_{n,l} \phi_{n,l}(z_1)\phi_{n,-l}(z_2),
\end{equation}
where $\phi_{n,l}$ are the NOs and $|c_{nl}|=|\sigma_{nl}|$ are the corresponding NAs. Then, the correlation coefficient takes the form \cite{PhysRev.172.49}
\begin{equation}\label{eq:tau_NO}
\tau = \frac{\sum_{n_1,n_2,l_1,l_2}c_{n_1,l_1} c_{n_2,l_2} |\bra{\phi_{n_1,l_1}}z\ket{\phi_{n_2,l_2}}|^2}{\sum_{n,l}(c_{n,l})^2\bra{\phi_{n,l}}z\ket{\phi_{n,l}}}.
\end{equation}
Note that only those terms for which $|l_1-l_2|=1$ contribute to the numerator; otherwise $\bra{\phi_{n_1,l_1}}z\ket{\phi_{n_2,l_2}}$ vanishes. Although the closed forms of the expectation values in the formula \eqref{eq:tau_NO} are difficult to find, numerics using the approximate NOs \eqref{eq:NO_asymptotics} shows that the convergence of the expression \eqref{eq:tau_NO} is rather slow, requiring the calculation of thousands of terms to get numerically close to the analytic value \eqref{eq:tau_final}.

\section{Discussion and Conclusions}

In this work we have analysed the asymptotic behaviour of the natural orbitals and their occupation numbers/natural amplitudes in the ground state of two non-interacting anyons in the boson magnetic gauge. We have derived exact asymptotic forms of the natural orbitals from a fixed $l$-sector of the 1RDM when the ordinal number of the orbital $n$ that indexes NOs and NAs tends to infinity (the latter are arranged descendingly according to their absolute values). While the asymptotic forms of NOs and NAs differ from their actual values for small $n$, the convergence is surprisingly fast when increasing $n$. 

Although our calculations were done for the ground state only, the methodology can be applied {\textit{mutatis mutandis}} to any other eigenstate (see \cite{wu1984harmonic} for their explicit forms) of this two-anyon system, resulting in the same NA asymptotic with suitably altered constant $\mathcal{D}(\alpha)$. Our asymptotic results may also extend to eigenstates of anyon gases with higher numbers of particles, however proving this would require using a different set of mathematical tools such as the ones applied in the work \cite{sobolev}.

The same method can be employed to derive the NO- and NA- asymptotic for the non-interacting two-anyon system at hand in the fermion magnetic gauge. The fermionic counterpart of the wavefunction \eqref{psiB} is \cite{Douglas}
\[\Psi_{F}^{(\alpha)} (z_1,z_2)  = \mathcal{N}_{\alpha}\, e^{i\theta_{12}}|z_1-z_2|^{\alpha} e^{-\frac{|z_1|^2+|z_2|^2}{2}},\]
where $\theta_{12}$ is the angle between $z_1$ and $z_2$. Note that the coalescence cusp in $\Psi_F$ is of the same order as the coalescence cusp of $\Psi_B$. By repeating the steps form Section \ref{sec:asymptotics} with $\Psi_B$ replaced by $\Psi_F$ we arrive at the identical conclusion concerning the asymptotics of NOs and NAs in the fermion magnetic gauge.

It is interesting to note that according to the power law\ \eqref{eq:NA_asymptotics}, the anyonic NONs in 2D decay slower than for Coulombic multi-electron systems in 3D. This confirms the intuition that 2D anyon systems are characterised by strong correlations and, in light of work \cite{C22}, are likely to be more challenging to tackle by the standard quantum chemistry toolset.

One might be tempted to interpret our presented results in terms of the Pauli exclusion principle for anyons. However, note that the magnetic gauge transformation \eqref{eq:statisticstrans} from the bosonic wavefunction $\Psi_B$ to the anyonyonic wavefunction $\Psi_\alpha$ is nonlocal, thus the NONs of $\Psi_B$ are different from the NONs of $\Psi_\alpha$. Moreover, only the NONs of $\Psi_\alpha$ are the ones that interpolate between fermionic case (Slater determinant for $\alpha=1$) and the bosonic case (fully condensed state for $\alpha=0$). Thus, for anyonic systems it is more appropriate to refer to gauge-invariant methods of measuring the Pauli exclusion principle such as the expectation value of the kinetic energy operator \cite{LS13a,LS13b,LS14}.

\begin{acknowledgments}
The authors would like to thank Jonathan Robbins for helpful discussions. The numerical computations presented here were conducted using the University of Bristol HPC system ({\textit{BlueCrystal 4}}). The research described in this publication has been funded by the National Science Center (Poland) under grant  2022/47/B/ST4/00002. The support from Max-Planck-Institut für Physik komplexer Systeme (MPI PKS), Dresden is also acknowledged by one of the authors (J.C.).
\end{acknowledgments}

\appendix
\section{The leading-order expansion of Equation \eqref{eq:integral_real_F}}\label{appendixA}
The paper \cite{kangshao2014} provides the following theorem:
for any $f$ being an analytic function in the region $\Omega=\{z\in\mathcal{C}:a\leq \Re(z)\leq b,\quad \Im(z)\geq 0\}$ we have
\begin{multline}\label{thm:integral}
     \int_a^b dx\, (x-a)^\alpha (b-x)^\beta f(x) e^{i\omega x}  \\
     = \frac{i^{\alpha+1}}{\omega^{\alpha+1}}e^{i\omega a}\,\int_0^\infty dp\, \left(b-a-i\frac{p}{\omega}\right)^\beta f\left(a+i\frac{p}{\omega}\right)p^\alpha e^{-p} \\
    -\frac{i^{\beta-1}}{(-1)^{\beta-1}\omega^{\beta+1}}e^{i\omega b}\,\int_0^\infty dp\, \left(b-a+i\frac{p}{\omega}\right)^\alpha \times \\
     \times f\left(b+i\frac{p}{\omega}\right)p^\beta e^{-p}.
\end{multline}
Theorem \ref{thm:integral} as stated originally in \cite{kangshao2014} has a typographic error and above we have provided its corrected version. In order to apply this theorem, we translate Equation\ \eqref{eq:integral_real_F} to the form
\begin{multline} \label{eq:fredholm_integral_sum}
    \mathcal{I}_-(u_1)+\mathcal{I}_+(u_1)=\\
    = \Re\left\{\int_0^{u_1} du_2\,F(u_2)\, \left(u_1-u_2\right)^{\alpha+1}e^{i\kappa_{nl} u_2}\right\} \\
    + \Re\left\{\int_{u_1}^{u_\infty} du_2\,F(u_2)\,\left(u_2-u_1\right)^{\alpha+1}e^{i\kappa_{nl} u_2}\right\}.
\end{multline}
Applying Theorem \ref{thm:integral} to the $\mathcal{I}_-(u_1)$,
\begin{multline*}
    \int_0^{u_1} du_2\, F (u_2) \left(u_1-u_2\right)^{\alpha+1}e^{i\kappa_{nl} u_2} \\
    = \frac{i}{\kappa_{nl}}\int_0^\infty dp\, \left(u_1 - i\frac{p}{\kappa_{nl}}\right)^{\alpha+1} F\left(i\frac{p}{\kappa_{nl}}\right) e^{-p} \\
    -\frac{i^{\alpha}e^{i\kappa_{nl} u_1}}{(-1)^{\alpha}{\kappa_{nl}}^{\alpha+2}}\int_0^\infty dp\, F \left(u_1 + i\frac{p}{\kappa_{nl}}\right)p^{\alpha+1} e^{-p}.
\end{multline*}
Extracting the leading order asymptotics in $\kappa_{nl}\to\infty$ produces
\begin{multline*}
    \frac{i}{\kappa_{nl}}\int_0^\infty dp\, u_1^{\alpha+1} F\left(0\right) e^{-p} \\
  -  \frac{i^{\alpha}e^{i\kappa_{nl} u_1}}{(-1)^{\alpha}{\kappa_{nl}}^{\alpha+2}}\int_0^\infty dp\, F \left(u_1\right)p^{\alpha+1} e^{-p} \\
    = \frac{i}{\kappa_{nl}} u_{1}^{\alpha+1} F (0) - \frac{i^{\alpha}e^{i\kappa_{nl} u_1}}{(-1)^{\alpha}{\kappa_{nl}}^{\alpha+2}} F \left(u_1\right) \Gamma(\alpha + 2),
\end{multline*}
for which it is straightforward to see that $F (r) \to 0$ as $r \to 0$.
Similarly, for the second integral we find
\begin{multline*}
    \int_{u_1}^{u_\infty} du_2\, F (u_2) \left(u_1-u_2\right)^{\alpha+1}e^{i\kappa_{nl} u_2} =\\
    = - \frac{i^{\alpha}}{\kappa_{nl}^{\alpha+2}} e^{i \kappa_{nl} u_1} F\left(u_1\right) \Gamma{(\alpha + 2)},
\end{multline*}
where we use the fact that $F (r)$ must tend to zero as $r \to \infty$. Summing up, the integral sum \eqref{eq:fredholm_integral_sum} can now be expressed as:
\begin{multline*}
    \mathcal{I}_-(u_1)+\mathcal{I}_+(u_1) = \nonumber \\
    = \Re \left\{ - i^{\alpha} \left( (-1)^{-\alpha} + 1 \right) \frac{e^{i \kappa_{nl} u_1}}{\kappa_{nl}^{\alpha+2}} F\left(u_1\right) \Gamma{(\alpha + 2)} \right\}\\
    =- 2 \cos\left(\frac{\pi\alpha}{2}\right)\, \cos(\kappa_{nl} u_1) \frac{F\left(u_1\right) \Gamma{(\alpha + 2)}}{\kappa_{nl}^{\alpha+2}} \nonumber
\end{multline*}
Using the above fact in \eqref{eq:fredholm_integral_sum} yields the Equation \eqref{eq:fredholm_asymp_reduced}.
\vfill
\bibliography{references.bib}

\end{document}